\documentclass[published]{JHEP}
\JHEP{01(2001)015}

\usepackage{amssymb}

\newcommand{\SO}{\mathop{\rm SO}}
\newcommand{\fr}[2]{{\frac{#1}{#2}}}
\newcommand{\dH}[2]{H^{(-)#1}_{#2}}
\newcommand{\uH}[2]{H^{(-)#1#2}}

\def\ra{\rightarrow}
\def\half{\frac{1}{2}}
\def\su{\sqrt{-u^2}}

\def\a{\alpha}
\def\b{\beta}
\def\d{\delta}

\def\e{\epsilon}
\def\ve{\varepsilon}
\def\f{\phi}
\def\F{\Phi}

\def\P{\Psi}

\def\L{\Lambda}
\def\m{\mu}
\def\n{\nu}
\def\r{\rho}
\def\s{\sigma}

\def\o{\omega}

\def\cD{{\cal D}}
\def\cF{{\cal F}}
\def\cL{{\cal L}}

\title{Antifield BRST quantization of duality-symmetric Maxwell theory}

\author{Xavier Bekaert\thanks{Work supported in part by
	the ``Actions de Recherche Concert{\'e}es" of the ``Direction de la
	Recherche Scientifique - Communaut{\'e} Fran{\c c}aise de Belgique",
	by IISN - Belgium (convention 4.4505.86).}  \\
	Centre \'Emile Borel, Institut Henri Poincar\'e \\ 
	11, rue Pierre et Marie Curie, 75231 Paris Cedex 05, France, and \\
	Physique Th\'eorique et Math\'ematique, Universit\'e Libre de Bruxelles \\
	Campus Plaine C.P. 231, B-1050 Bruxelles, Belgium \\
	E-mail: \email{xbekaert@ulb.ac.be}} 

\author{Sorin Cucu\thanks{Work supported by the European Commission TMR
	programme HPRN-CT-2000-00131, in which X.B. is associated to Leuven.} \\
	Centre \'Emile Borel, Institut Henri Poincar\'e \\ 
	11, rue Pierre et Marie Curie, 75231 Paris Cedex 05, France, and \\
	Instituut voor Theoretische Fysica, Katholieke Universiteit Leuven \\ 
	Celestijnenlaan 200D, B-3001 Leuven, Belgium \\
	E-mail: \email{sorin.cucu@fys.kuleuven.ac.be}}

\abstract{We perform the antifield BRST quantization of
duality-symmetric Maxwell theory and show explicitly the quantum
equivalence of the different formulations (covariant and
non-covariant).  The non-covariant gauge-fixed action is used in the
computation of propagators for this model.}

\received{December 6, 2000}
\accepted{January 12, 2001}


\begin{document}

\section{Introduction}

In the past few years the concept of duality played a central role in
field and string theory. Dualities became systematically studied in
the literature once their importance in connecting apparently
different string theories was realized. For instance S-duality
establishes a correspondence between weak and strong coupled models
and a special case of S-duality is represented by the
electric-magnetic duality. Thus, the necessity of studying such a
duality required a dual-symmetric action for the Maxwell theory.  This
manifest duality symmetry can be elegantly reformulated in terms of a
self-duality condition on a complex field strength.  Abelian
$p$-forms, with $(p+1)$-field strengths satisfying a self-duality
condition (Hodge duality), are only defined in $2(p+1)$ dimensions.
Because of the minkowskian signature, the square of the Hodge dual $*$
is the identity in twice odd dimensions and minus the identity in
twice even dimensions Thus, the condition: $F=*F$ allows non-trivial
solutions ($F\neq 0$) for real fields only in twice odd dimensions:
the chiral $p$-forms.  In twice even dimensions, we have to take $F$
to be a complex field and redefine the dual operator to be imaginary
by $*\rightarrow {\rm i}*$.  The complexification of the fields is
also equivalent to the dualization of a pair of real $p$-forms gauge
fields, like duality-symmetric Maxwell theory ($p=1$).

The connection of chiral $p$-forms to supergravity~\cite{sw-hw} or
branes and M-theory~\cite{bb-ckvp} was one of the motivations for a
methodical approach of the subject. Several non-covariant
actions~\cite{fj,Henneaux:1988,ms} were proposed for the description
of chiral $p$-forms. The main obstacle encountered in the construction
of an action with manifest Lorentz-invariance was the presence of the
self-duality requirement. Nevertheless, the problem was solved either
by introduction of an infinite set of auxiliary fields entering the
lagrangian in a polynomial way~\cite{wy,dh} or using one auxiliary
field in a non-polynomial way~\cite{pst1,mps}.  Efforts for
implementing the duality symmetry in Maxwell theory at the level of
its action have been undertaken since the
seventies~\cite{z:71,dt}. The topic has been addressed again over the
last decade in a series of papers~\cite{ss,pst}. This led to
non-covariant versions~\cite{dt,ss} or to lagrangians with manifest
space-time symmetry~\cite{pst,mb}.

The quantization of theories containing chiral $p$-forms has been
already performed for several values of $p$ and different formulations
of the systems. The covariant hamiltonian BRST
(Becchi-Rouet-Stora-Tyutin) quantization of one chiral boson was
realized in~\cite{wy} and generalized to chiral $p$-forms
in~\cite{dh}, applying the formulation of infinitely many ghosts. On
the other hand, chiral $2$--forms in $6$ dimensions have been recently
quantized~\cite{kor} within the covariant BV (Batalin-Vilkovisky)
treatment making use of various gauge-fixing conditions. The BV method
has been also adopted~\cite{ggrs} in proving the quantum equivalence
of Schwarz-Sen~\cite{ss} and Maxwell~\cite{dt} theories. Nevertheless,
the generating functionals derived in~\cite{ggrs} do not exhibit a
manifest Lorentz covariance.  The aim of the present work is to obtain
a correct path-integral for the covariant duality-symmetric Maxwell
theory. As, in the first instance, we want to get a generating
functional with manifest Lorentz symmetry we will base our
considerations on the action proposed by Pasti, Sorokin and Tonin
(PST) in~\cite{pst}. The presence of the auxiliary field coupling
(non-polynomially) to the two gauge potentials makes the gauge algebra
non-Abelian, with field-dependent structure \emph{functions}. As a
consequence, we must choose a suitable quantization procedure. We will
consider here the antifield-BRST method~\cite{bv} because it proved,
in the last twenty years, to be a very powerful quantization technique
applicable also for models with open and/or non-Abelian
(field-dependent) algebras, as it will be the case for us.

The paper is structured as follows. In section~\ref{s:action} we
present the action and its gauge symmetries together with the gauge
algebra. We compute then, in section~\ref{s:BRST}, the minimal
solution of the master equation and we infer also the BRST
symmetry. By a well chosen non-minimal sector and an adequate
gauge-fixing fermion the remaining gauge invariances will be fixed in
section~\ref{s:g-fixing}. The non-covariant gauge is the starting
point in proving the quantum equivalence between the PST and the
Maxwell theories as explained in section~\ref{s:qmequiv}. It is
afterwards used to explicitly determine the Feynman rules for the
interaction of PST with gravity in section~\ref{s:gravity}. In the
last part, section~\ref{s:concl}, we collect and discuss our results.

\section{Gauge symmetries of the classical action}\label{s:action}

We start our discussion by considering the PST action proposed to
manifestly implement two symmetries in the description of free Maxwell
theory: Lorentz invariance and electric-magnetic duality.

After fixing the notation, we emphasize the physical content of this
model. Next, we briefly present its gauge algebra.

The PST action~\cite{pst} constructed for the description of self-dual
vector field is
\begin{equation}
S_0=\int d^4x \left(-{1\over 8}F^\alpha_{mn}F^{\alpha mn}
+{1\over{4(-u_lu^l)}}u^m{\cal F}^\alpha_{mn}{\cal F}^{\alpha np}u_p
\right),
\label{e:pst}
\end{equation}
where the $m,n,\dots\,$ stand for Lorentz indices in $4$ dimensional
space-time with a flat metric $(-,+,+,+)$. As explained in the
introduction the Lagrangian contains two gauge potentials
$(A_m^\a)_{\a=1,2}$ and one auxiliary field $a$, appearing here only
as the gradient $u_m=\partial_m a$.  The notation used throughout this
paper is
\begin{equation}
\begin{array}[b]{rclcrcl}
u^2&=&u^mu_m \,,
&\qquad&
v_m&=&\displaystyle \fr{u_m}{\su}\,,
\\[8pt]
F^\a_{mn}&=&2\partial_{[m}A^\a_{n]}\,,
&\qquad&
 F^{*\a}_{mn}&=&\displaystyle \half\e_{mnpq}F^{\a pq} \,,
\\ 
\cF^\a_{mn}&=&\cL^{\a\b}F^\b_{mn}-F^{*\a}_{mn}\,,
&\qquad&
\dH{\a}{m}&=&\cF^\a_{mn}v^n  
\end{array}
\label{e:not}
\end{equation}
with $\cL^{\a\b}$ being the antisymmetric unit matrix of $\SO(2)$.
The equations of motion associated to~(\ref{e:pst}) read
\begin{eqnarray}
\d A_m^\a &:\qquad &  \e^{mnqp}\partial_n(v_pH^{(-)\a}_q)=0\,,
\label{e:eomA}\\
\d u_m & :\qquad & \fr{1}{2\sqrt{-u^2}}\left(H^{(-)\a}_n
\cF^{\a mn}-H^{(-)\a}_nH^{(-)\a n}v^m\right)=0\,.
\label{e:eomu}
\end{eqnarray}
It is straightforward to check the following gauge invariances
of~(\ref{e:pst})
\begin{eqnarray}
\d_IA_m^\a&=&\partial_m\varphi^\a\,, \qquad \d_I a=0 \,,
\label{e:ginv1}\\
\d_{II}A_m^\a&=&-\cL^{\a\b}\dH{\b}{m}\fr{\f}{\sqrt{-u^2}}\,, \qquad \d_{II}a=\f \,,
\label{e:ginv2}\\
\d_{III}A_m^\a&=&u_m\ve^\a\,, \qquad \d_{III}a=0 \,,
\label{e:ginv3}
\end{eqnarray}

\pagebreak[3]

\noindent that are irreducible.  Pasti-Sorokin-Tonin have shown~\cite{pst} that
this model is in fact classically equivalent with Schwarz-Sen
action~\cite{ss} describing the dynamics of a single Maxwell
field. Indeed, using the equations of motion~(\ref{e:eomA}) one can
fix the gauge degrees of freedom of~(\ref{e:ginv3}) in such a way that
the self-duality condition
\begin{equation}
\cF^{\a}_{mn}=0\label{e:duality}
\end{equation}
is satisfied. Such a consequence of the equations of motion allows us
to express one of the gauge fields $A_m^\a$ as function of the other
one yielding the usual Maxwell Lagrangian (with remaining
symmetry~(\ref{e:ginv1})) plus a contribution of $u_m$ field. Further,
one remarks from the second invariance~(\ref{e:ginv2}) that $a$ is
pure gauge. Another way to see that is by expressing the field
equation for $a$ as a consequence of the equation of motion for
$A_m^\a$. That is why $a$ can be easily fixed away using a clever
gauge condition (avoiding the singularity $u^2=0$). So, the field
$u_m$ as well as one of the two $A_m^\a$ are auxiliary in the sence
that one needs them only to lift self-duality and Lorentz invariance
at the rank of manifest symmetries of the action. But, they can be
removed on the mass-shell taking into account the gauge invariances of
the new system. Nevertheless, the way we gauge fix the last
invariance~(\ref{e:ginv3}) can be applied only at the classical level
since we make explicit use of the field equations, which cannot be
done in a BRST path integral approach. The manner of fixing the
unphysical degrees of freedom in the BRST formalism will be clarified
in section~\ref{s:g-fixing}.

Computing the gauge algebra we get
\begin{eqnarray}
[\d_{II}(\f_1),\d_{II}(\f_2)]&=&\d_{III}\left(\fr{\cL^{\a\b}\dH{\b}{p}}
{(-u^2)^{3/2}}(\f_1\partial^p\f_2-\f_2\partial^p\f_1)\right)\,,
\label{e:galg1}\\
\left[\d_{II}(\f),\d_{III}(\ve^\a)\right]&=&\d_I(\f\ve^\a)+\d_{III}
\left(\fr{u^p\f}{(-u^2)} \partial_p\ve^\a\right)\,.
\label{e:galg2}
\end{eqnarray}
Thus, our system describes a non-Abelian theory with the structure
constants replaced by non-polynomial structure \emph{functions}.

\section{Minimal solution of the master equation}\label{s:BRST}

Having made the classical analysis of the model, we can start now the
standard BRST procedure.\footnote{For a short review of antifield BRST
method see appendix~\ref{a:BRST}.} The first step is to construct the
minimal solution of the master equation with the help of the gauge
algebra. In order to reach that end we will introduce some new fields
called ghosts and their antibracket conjugates known as antifields.

The minimal sector of fields and antifields dictated by the gauge
invariances~(\ref{e:ginv1}) --(\ref{e:ginv3}) as well as their ghost
numbers and statistics are listed in table~\ref{t:mingh}.

\TABULAR{|c|c|c|c|c|c|c|c|c|c|c|}{\hline
$\Phi$&$A_m^\a$&$a$&$A_m^{\a *}$&$a^*$&$c^\a$&$c$&$c'^\a$&$c^{\a *}$&$c^*$&$c'^{\a *}$\\\hline
$gh(\Phi)$&$0$&$0$&$-1$&$-1$&$1$&$1$&$1$&$-2$&$-2$&$-2$\\\hline
${\rm antigh}(\Phi)$&$0$&$0$&$1$&$1$&$0$&$0$&$0$&$2$&$2$&$2$\\ \hline
\mbox{stat($\Phi$)}&$+$&$+$&$-$&$-$&$-$&$-$&$-$&$+$&$+$&$+$\\\hline}%
{Ghost number, antighost number and statistics of the minimal fields and their 
antifields.\label{t:mingh}}

The transformations~(\ref{e:ginv1})--(\ref{e:ginv3}) determine directly
the antigh number one piece of the extended action, i.e.
\begin{equation}
S_1=\int d^4\,x\left[A_m^{\a *}\left(\partial^mc^\a- \cL^{\a\b}\dH{\b}{m}
\fr{c}{\sqrt{-u^2}}+u^mc'^\a\right)+ a^* c\right].
\label{e:ext1}
\end{equation}
In order to take into account the structure functions one has to
insert in the solution of the master equation a contribution with
antigh number two of the form
\begin{equation}
\label{e:ext2}
S_2 = \int d^4\,x \left[c'^{\a *}\left( \fr{\cL^{\a\b}\dH{\b}{p}}
{(-u^2)^{3/2}}c\,\partial^pc +\fr{v_p}{\su}c\partial^p c'^{\a}\right) +c^{\a *}c\,c'^\a \right].
\end{equation} 
Due to the field dependence of the structure functions one should
expect that $S_1$ and $S_2$ are not enough to completely determine the
extended action and one will need an extra piece of antigh number
three to do the job. Indeed, that was already the case for chiral
2--forms in 6 dimensions discussed in~\cite{kor}. Nevertheless, one
can readily check that in the present situation $S_{\rm
min}=S_0+S_1+S_2$ is the minimal solution of the classical master
equation $(S_{\rm min},S_{\rm min})=0$, i.e.
\begin{eqnarray}
(S_1,S_1)_1 + 2(S_1,S_1)_1 &=&0\,,
\nonumber \\
(S_2,S_2)_2 + 2(S_1,S_2)_2 &=&0\,.
\end{eqnarray}
This follows also as a consequence of the irreducibility of our
model. The way we constructed $S_{\rm min}$ will ensure also a
properness condition which says that the rank of the hessian
\begin{equation}
S_{\tilde C}{}^{\tilde A} = \o^{{\tilde A}{\tilde B}}
\frac{\d^L\d^R S_{\rm min}}{\d \Phi^{\tilde B}\d \Phi^{\tilde C}}
\end{equation}
at the stationary surface corresponds to precisely to half its
dimension (where \linebreak$(\Phi^{\tilde A})_{{\tilde A}=1,\dots
,2N}$ labels all the fields and antifields in the minimal sector,
while $\o^{{\tilde A}{\tilde B}}$ denotes the symplectic matrix in
$2N$ dimensions). Such a condition expresses that $S_{\rm min}$ has
only a number of gauge invariances equal to $N$, not $2N$ as one could
superficially think.

Once $S_{\rm min}$ has been derived, we can infer the BRST operator
$s$, which is the sum of three operator of different antigh
number
\begin{equation}
s=\delta+\gamma+\rho\,.
\end{equation}
For instance, the non-trivial action of the Koszul-Tate differential,
of antigh number $-1$, is in our case given by
\begin{eqnarray}
\d A_m^{\a *} &=& \e^{mnqp}\partial_n(v_pH^{(-)\a}_q)\,,\\
\d a^* &=& \partial_m\left(\fr{1}{2\sqrt{-u^2}}\left(H^{(-)\a}_n\cF^{\a mn}
-H^{(-)\a}_nH^{(-)\a n}v^m\right)\right)\,,\\
\d c^{\a *}&=&-\partial^m A^{\a *}_m\,,
\label{e:KT1} \\
\d c^* &=&-\fr{\cL^{\a\b}\uH{\b}{p}}{\su} A^{\a *}_p +a^*\,,
\label{e:KT2} \\
\d c'^{\a *}&=&u^m A^{\a *}_m\,.
\label{e:KT3}
\end{eqnarray}
The third piece, $\rho$, of antigh number $+1$ is present also
because the structure functions determined
by~(\ref{e:ginv1})-(\ref{e:ginv3}) depend explicitly on the fields.

In this way the goal of this section, i.e.\ the construction of the
minimal solution for the master equation, was achieved.

\section{The gauge-fixed action}\label{s:g-fixing}

The minimal solution $S_{\rm min}$ will not suffice in fixing all the
gauge invariances of the system and, before fixing the gauge, one
needs a non-minimal solution for $(S,S)=0$ in order to take into
account the trivial gauge transformations.  In this section we first
construct such a non-minimal solution and, afterwards, we propose two
possible gauge-fixing conditions which will yield two versions for the
gauged-fixed action: a covariant and a non-covariant one.

\subsection{Non-minimal sector}

Inspired by the gauge transformations~(\ref{e:ginv1})-(\ref{e:ginv3})
and their irreducibility we propose a non-minimal sector given in
table~\ref{t:non-minsect}.

\TABULAR[b]{|c|c|c|c|c|c|c|c|c|c|c|c|c|}{ \hline
$\Phi$&$B^\a$&$B$&$B'^\a$&${\bar C}^\a$&${\bar C}$&${\bar C}'^\a$&
$B^{\a *}$&$B^*$&$B'^{\a *}$&${\bar C}^{\a *}$&${\bar C}^*$&${\bar
C}'^{\a *}$\\ \hline
$gh(\Phi)$&$0$&$0$&$0$&$-1$&$-1$&$-1$&$-1$&$-1$&$-1$&$0$&$0$&$0$\\\hline
\mbox{stat($\Phi$)}&$+$&$+$&$+$&$-$&$-$&$-$&$-$&$-$&$-$&$+$&$+$&$+$\\\hline}%
{Ghost number and statistics of the non-minimal fields and
their antifields.\label{t:non-minsect}}

They satisfy the following equations
\begin{eqnarray}
s{\bar C}^{\cdots} &=& B^{\cdots}\,,
\nonumber\\
\qquad sB^{\cdots}&=&0\,, 
\nonumber\\
sB^{\cdots *}&=&{\bar C}^{\cdots *}\,,
\nonumber\\
s{\bar C}^{\cdots *} &=&0\,.
\end{eqnarray}
The dots are there to express that these relations are valid for the
correspondingly three kinds of non-minimal fields.  We immediately see
that $\bar{C}^{\cdots}$'s and $B^{\cdots}$'s constitute trivial pairs,
as well as their respective antifields, in such a way that they do not
enter in the cohomology of $s$. Hence, they are called non-minimal. A
satisfactory explanation of the necessity of the presence of a
non-minimal sector is provided by BRST-anti-BRST formalism.  Their
contribution to the solution of the master equation is
\begin{equation}
\label{e:non-min}
S_{\rm non-min}=S_{\rm min} +\int d^4\,x \left({\bar C}^{\a *}B^\a +
{\bar C}^* B +{\bar C}'^{\a *}B'^\a \right).
\end{equation}

\subsection{Covariant gauge fixing}

We will first try a covariant gauge fixing that in principle should
yield a covariant gauge-fixed action and we will see what is the main
problem that occurs. One can consider the following \emph{covariant}
gauge choices
\begin{eqnarray}
\d_{I} &\ra & \partial^m A^\a _m =0\,,
\label{e:gc1} \\
\d_{II} &\ra & u^2 +1=0\,,
\label{e:gc2} \\
\d_{III} &\ra & u^m A^\a _m =0\,.
\label{e:gc3} 
\end{eqnarray}
The gauge choice~(\ref{e:gc1}) is analogous to the Lorentz gauge. In
its turn~(\ref{e:gc2}) allows to take a particular Lorentz frame in
which $u^m(x)$ is the unit time vector at the point $x$. In such a
case, at the point $x$,~(\ref{e:gc3}) is the temporal gauge condition
for the two potentials.

A gauge-fixing fermion corresponding to the gauge
choices~(\ref{e:gc1})-(\ref{e:gc3}) is
\begin{equation}
\label{e:gaugeferm}
\P[\F^A]=-\int d^4\,x \left[{\bar C}^\a \partial^m A^\a _m  
+{\bar C} (u^2 +1) +{\bar C}'^\a u^m A^\a _m \right]. 
\end{equation}
One expresses now all the antifields with the help of $\P[\F]$, i.e.
\begin{equation}
\label{e:antifields}
\F^*_A =\fr{\d\P[\F^A]}{\d\Phi^A}
\end{equation}
getting
\begin{eqnarray}
\label{e:antifield1}
A_m^{\a *} &=& \partial_m {\bar C}^\a -u_m {\bar C}'^\a\,, 
\qquad a^*= 2\partial_m (u^m {\bar C}) +\partial^m(A_m^\a {\bar C}'^\a) \,, 
\nonumber\\
\label{antifield2}
c^{\cdots *}&=& 0\,, \qquad B^{\cdots *}= 0\,, 
\nonumber\\
{\bar C}^{\a *} &=& -\partial^m A^\a _m \,, \qquad 
{\bar C}^* = -(u^2 +1) \qquad {\bar C}'^{\a *} = -u^m A^\a _m  \,.\label{e:antifield3}
\nonumber
\end{eqnarray}
Using the last relations one can find the gauge fixed action as
in~(\ref{e:GAUGEFIXED}) which in our case reads
\begin{eqnarray}
\label{e:fixedaction}
S_\P &=& S_0 + \int d^4\,x \Biggl[-{\bar C}^\a \square c^\a -
\fr{\cL^{\a\b}\uH{\b}{m}}{\sqrt{-u^2}}\partial_m {\bar C}^\a \cdot c
+u^m \partial_m {\bar C}^\a \cdot c'^\a - 
\nonumber\\&&
	\hphantom{S_0 + \int d^4\,x \Biggl[}\!
-u_m {\bar C}'^\a \partial^m c^\a - u^2{\bar C}'^\a c'^\a -(2 u_m
{\bar C} + A_m^\a {\bar C}'^\a)\partial^m c -
\nonumber\\ &&
	\hphantom{S_0 + \int d^4\,x \Biggl[}\!
-\,(\partial^m
A^\a _m )B^\a -(u^2+1)B -(u^m A^\a _m )B'^\a \Biggr]\,.
\end{eqnarray}
Writing down the path integral~(\ref{e:PATHINT}), one can integrate
directly the fields $B^{\cdots}$ producing the gauge
conditions~(\ref{e:gc1})-(\ref{e:gc3}). A further integration of
${\bar C}$, ${\bar C}'^\a$ and $c'^\a$ (in this order) leads to
\begin{equation}
\label{e:fixed1}
Z = \int\left[\cD A\cD a\cD c^\a \cD c\cD {\bar C}^\a \right] \,\d
(\partial^m A_m^\a) \,\d (u^2+1) \,\d (u^mA_m^\a) \,\d (u^m \partial_m
c) \exp iS'_\Psi\,,
\end{equation}
where
\begin{equation} 
\label{e:covgfa}
S'_\Psi =\int d^4x\left[ -{\bar C}^\a \square c^\a +\left(
\fr{\cL^{\a\b}\uH{\b}{m}}{\sqrt{-u^2}} c +\fr{u^m}{u^2} (u^p\partial_p
c^\a +A_p^\a\partial^p c )\right)\partial_m {\bar C}^\a\right].
\end{equation}
Of course, the next step in getting a covariant generating functional
from which we should read out the \emph{covariant} propagator for the
fields $A_m^\a$ would be the elimination of $c$ and $a$
in~(\ref{e:fixed1}). Due to the ``gauge condition" for the ghost $c$
(i.e.\ $u^m \partial_m c =0$) and the way it enters the gauge-fixed
action $S'_\Psi$, this integration is technically difficult. What one
could try is to integrate both $c$ and $a$ at the same time. This is
also not straightforwardly possible as a consequence of the gauge
condition~(\ref{e:gc2}). This requirement was necessary to
\emph{covariantly} fix the symmetry~(\ref{e:ginv2}). Nevertheless, one
can attempt to find the general solution to this
equation~(\ref{e:gc2}), which reduces to the integration of
$\partial^m a=\L^m{}_p(x)n^p $ (with $\L^m{}_p(x)$ a point-dependent
Lorentz boost and $n_p$ a constant time-like vector, i.e.\ $n_p n^p
=-1$). Such a solution is still unconvenient due to $x$-dependence of
the Lorentz transformation matrix $\L^m{}_p(x)$.

A way to overcome this sort of complication is to choose a particular
form for this matrix, breaking Lorentz symmetry. It is precisely this
price that we have to pay in order to explicitly derive the propagator
of $A_m^\a$ fields. As it will be explained in the next subsection, by
taking a particular solution for~(\ref{e:gc2}), i.e.\ by giving up
Lorentz invariance, we will be able to express the gauged-fixed action
in a more convenient form for our purposes.

\subsection{Non-covariant gauge fixing}

As it was remarked in the previous subsection in order to explicitly
derive the Feynman rules for the PST model one has to break up its
Lorentz symmetry by taking a specific solution of the
equation~(\ref{e:gc2}). In this subsection we present a non-covariant
gauge of the theory and the advantages for such a choice will become
clear in the next sections. A possible \emph{non-covariant} gauge
fixing is
\begin{eqnarray}
\d_{I} &\ra & \partial^m A^\a _m =0\,,
\label{e:gnc1} \\
\d_{II} &\ra & a-n_m x^m=0\,,\qquad n_m n^m =-1 \,,
\label{e:gnc2} \\
\d_{III} &\ra & n^m A^\a _m =0\,.
\label{e:gnc3} 
\end{eqnarray}
By~(\ref{e:gnc2}), the gradient $\partial^m a$ becomes equal to the
vector $n^m$ introduced above.  In a Lorentz frame where
$n^m=(1,0,0,0)$ the requirement~(\ref{e:gnc3}) is the temporal gauge
condition and~(\ref{e:gnc1}) the Coulomb gauge condition for the two
potentials $A^\a_m$.

Then, the gauge-fixing fermion will be
\begin{equation}
\label{e:gaugeferm-non}
\P[\F^A]=-\int d^4\,x \left[{\bar C}^\a \partial^m A^\a _m +{\bar C}
(a-n_m x^m) +{\bar C}'^\a n^m A^\a _m \right].
\end{equation}

Using the same non-minimal contribution $S_{non-min}$ as before, the
non-covariant gauge-fixed action is
\begin{eqnarray}
\label{e:gauge-fixedaction}
S_\P &=& S_0 + \int d^4\,x \Biggl[\left(\partial_m {\bar C}^\a -u_m {\bar
C}'^\a \right)\left(\partial^mc^\a-
\cL^{\a\b}\dH{\b}{m}\fr{c}{\sqrt{-u^2}}+u^mc'^\a\right) +
\nonumber\\ &&
	\hphantom{S_0 + \int d^4\,x \Biggl[}\!
+\left(-{\bar C}+\partial^m(A_m^\a {\bar C}'^\a)\right) c-\partial^m
A^\a _mB^\a - (a-n_m x^m)B -\,
\nonumber\\ &&
	\hphantom{S_0 + \int d^4\,x \Biggl[}\!
-u^m A^\a _mB'^\a \Biggr]\,.
\end{eqnarray}
This action is by far more convenient in deriving the propagator of
the gauge fields than its covariant expression~(\ref{e:covgfa})
because one can completely integrate the ghost sector. Also, the
bosonic part takes a more familiar form.  The quantum equivalence of
the PST model with ordinary Maxwell theory will be based also on this
non-covariant action.

\section{Path integral, quantum equivalence of PST action with Maxwell theory}\label{s:qmequiv}

The gauge-fixed action corresponding to the non-covariant gauge choice
can be used to recover the Schwarz-Sen theory, which is itself
equivalent to the Maxwell theory.  The generating functional is taken
to be (see appendix~\ref{a:BRST} and~\ref{a:Max})
\begin{equation}
Z= \int {\cal D} A^\a_m \,\cD a \,\cD c^{\cdots} \,\cD B^{\cdots}
\,\cD \bar{C}^{\cdots} \, {\rm det}(\square)\,{\rm det} ^{-1}({\rm
curl})\, \exp iS_\P\,,
\end{equation}
where $S_\P$ is given by (\ref{e:gauge-fixedaction}).

After integrating out some fields, in the following order
($B^{\cdots},$ $\bar{C},$ $c,$ $\bar{C}'^\a,$ $c'^\a ,$ $a$), we
obtain the path integral
\begin{equation}
Z= \int {\cal D} A^\a_m \,\cD {\bar C}^\a\,\cD c^\a \,{\rm
det}(\square)\, \,{\rm det} ^{-1}({\rm curl})\,\delta(\partial^m
A^\a_m)\,\delta(n^m A^\a_m)\, \exp iS'_\P\,,
\end{equation}
where the gauge-fixed action reduces now to
\begin{equation}
\label{e:non-fixedaction1}
S'_\P =\int d^4\,x \left[ -\fr12 n^mF^{*\a}_{mn}\cF^{\a np}n_p -{\bar
C}^\a \square c^\a -{\bar C}^\a n^pn^q\partial_p\partial_q c^\a \right].
\end{equation}
If we place ourselves in a Lorentz frame where $n^m=(1,0,0,0)$, the
functional $S'_\P$ assumes the form of the sum of Schwarz-Sen
gauge-fixed action~(\ref{e:ScSe}) and a ghost term
\begin{equation}
\label{e:ghosts}
-\int d^4 x\, {\bar C}^\a \bigtriangleup c^\a \,. 
\end{equation}
At this point we can integrate the ghosts $\bar{C}^\a$ and $c^\a$, and
the two fields $A^\a_0$, obtaining exactly the generating
functional~(\ref{e:genfunc}) of the Maxwell theory in the
non-covariant formulation (The quantum equivalence of Maxwell and
Schwarz-Sen actions is briefly reviewed in appendix~\ref{a:Max}).

This proves the (quantum) equivalence of the PST action~(\ref{e:pst})
with the Maxwell theory, which was already known at the classical
level.  The quantum equivalence was not obvious because the PST action
of the free Maxwell theory is not quadratic (and so the path integral
is not gaussian) and the pure gauge field $a$ is not, strictly
speaking, an auxiliary field (its equation of motion is not an
\emph{algebraic} relation which allows its elimination from the
action).

As a last remark, we notice that the bosonic part of the
action~(\ref{e:non-fixedaction1}) produces two poles in the propagator
of the gauge fields $A_i^\a$ : one physical, the usual
${1}/{\square}$, the other is an \emph{apparent} unphysical pole of
type $-{1}/({\partial^2_N +\square})$ (i.e.\ the inverse of laplacian
operator in an appropriate Lorentz frame, as $\partial_N \equiv
n^p\partial_p$), also present for the Schwarz-Sen theory. This is not
a physical pole because it corresponds to modes that do not
propagate. This pole appears also in another non-covariant gauge
choice: the Coulomb gauge in Maxwell theory. We will try to clarify
this point in the next section using the example of PST model coupled
to gravity. In that case, we can see explicitly that the unphysical
mode do not contribute at all to scattering amplitudes.

\section{Coupling to gravity} \label{s:gravity}

The goal of this section is to show that the massless propagator
${1}/{\square}$, rather than $-{1}/({\partial^2_N +\square})$,
contributes to the Feynman diagrams of the vector fields $A_m^\a$ for
the particular case of PST theory coupled to gravity.

We consider now the same PST action~(\ref{e:pst}) but in a
gravitational background characterized by a metric $g_{\mu\nu}$,
i.e.\ we take\footnote{The greek letters $\m$, $\n$, $\r$ etc. label
curved indices, while $\a$ and $\b$ denote $\SO(2)$ indices.}
\begin{equation}
\label{e:pstgrav}
S^g_0=-\frac{1}{2}\int \sqrt{-g} \, v_\mu
\dH{\a}{\nu}g^{\m\r}g^{\n\sigma}F^{*\a}_{\r\sigma}\,,
\end{equation}
where $g=\det{g_{\m\n}}$.

Next, we apply the same BRST formalism as before, following precisely
the same steps (just replacing the flat indices by curved
ones). Moreover, using the same non-covariant gauge, one infers for
the ghost sector a similar contribution of type~(\ref{e:ghosts}),
which becomes here
\begin{equation}
-\int d^4\,x \sqrt{-g}\,{\bar C}^\a (\square_{cov}
 +\nabla^{cov}_N\nabla^{cov}_N )c^\a \,.
\end{equation}
The only difference resides in replacing the ordinary derivatives by
covariant quantities. In any case, the fermionic ghosts decouple from
the bosonic fields and can be handled as explained
after~(\ref{e:ghosts}) (by integrating over them in the path
integral). This is the reason way we focus our attention on the
bosonic part of the gauge-fixed action arising from the original
action $S^g_0$.

As we are looking only for the first-order interaction of the model
with the background $g_{\m\n}$ it is natural to try to expand this
metric around the flat one. In other words, we consider
\begin{equation}
g^{\m\n}=\eta^{\m\n} + h^{\m\n}
\end{equation}
and, for further convenience, we assume that the fluctuation
$h^{\m\n}$ can be parametrized in terms of inverse $e_m^\m$ of the
orthogonal vectors $e_\m^m$, i.e.
\begin{equation}
h^{\m\n} = e_m^\m  e_n^\n \eta^{mn}
\end{equation}
(for simplicity the label $m$ will be suppressed in the future considerations). 

Our next move consists in developing the bosonic part of the
gauge-fixed action to the first-order in the perturbation $h^{\m\n}\,$.
\footnote{The indices are from now on raised and lowered with the flat
metric $\eta^{\m\n}$.} After some computation one gets
\begin{eqnarray}
\label{e:order0}
S^g_\P &=& \int d^4\,x \Biggl\{-\frac{1}{2}
A_\m^\a[\d^{\a\b}\eta^{\m\n}(-\square-\partial^2_N)
+\cL^{\a\b}T^{\m\n}\partial_N ]A_\n^\b + 
\\&&
	\hphantom{\int d^4\,x \Biggl\{}\!
+ \frac{1}{2} (T^{\m\n}A^\a_\n)\Biggl[\d^{\a\b}\eta_{\m\s}(\frac{1}{2}{\tilde
h}-(n^\tau e_\tau)^2) +\d^{\a\b}e_\m e_\s - 
\label{e:1order1}\\&&
	\hphantom{+ \frac{1}{2} (T^{\m\n}A^\a_\n)\Biggl[}\!
-\frac{1}{2} \cL^{\a\b}(n^\zeta e_\zeta)\e_{\m\kappa\tau\s}e^\kappa
 n^\tau\Biggr](T^{\s\r}A^\b_\r)\Biggr\} \,,
\label{e:2order1}
\end{eqnarray}
where we neglected the second order in $h^{\m\n}$ or higher. In the
meantime we have employed the notation ${\tilde h} =
h^{\m\n}\eta_{\m\n}$ and
\begin{equation}
\label{e:T}
T^{\m\n}A_\n^\a = \e^{\m\n\r\s}n_\r \partial_\s A_\n^\a \,.
\end{equation}
The object $T^{\m\n}$, defined in this way, is a differential operator
transforming one-forms into one-forms. It is antisymmetric under the
interchange of its indices and it is characterized by a very important
feature, namely
\begin{equation}
\label{e:propT}
T^{\m\r}T_{\r\s}T^{\s\n} = - (\square +\partial_N^2) T^{\m\n}\,.
\end{equation}
This property allows one to transform any series expansion in
$T^{\m\n}$ into a polynomial containing only $1$, $T$ and $T^2$.

Let us return to the interpretation of the
expansion~(\ref{e:order0})-(\ref{e:2order1}). The first remark is that
the zeroth-order,~(\ref{e:order0}), coincides with the one from the
flat space discussion.

This term delivers the gauge-fixed kinetic operator 
\begin{equation}
K_{\m\n}{}^{\a\b} = \d^{\a\b}\eta_{\m\n}(-\square-\partial^2_N)
+\cL^{\a\b}T_{\m\n}\partial_N\,,
\end{equation}
whose inverse is nothing but the propagator $P_{\m\n}{}^{\a\b}$ of the
vector fields $A_\m^\a$. Then a simple computation based on the
property~(\ref{e:propT}) of $T^{\m\n}$ gives the explicit form of the
propagator
\begin{equation}
\label{e:propag}
P_{\m\n}{}^{\a\b} = -\fr{1}{\square +\partial_N^2}\left[
\d_{\m\n}\d^{\a\b} + \fr{\cL^{\a\b}T_{\m\n}\partial_N}{\square}
-\fr{\d^{\a\b} T_{\m\r}T^\r{}_\n \partial_N^2}{\square(\square
+\partial_N^2)}\right].
\end{equation}

If we consider also the first-order
interaction~(\ref{e:1order1})-(\ref{e:2order1}) with a gravitational
background we notice that in such an interaction the gauge fields
$A_\m^\a$ couple to the perturbation $h^{\m\n}$ only as
$T^{\m\n}A_\n^\a$. Therefore, we conclude that the effective
propagator in the presence of gravity must be
\begin{equation}
\label{e:effpropag}
T^{\m\r}P_{\r\s}{}^{\a\b}T^{\s\n} = [\cL^{\a\b}\d^\m{}_\s \partial_N
-\d^{\a\b} T^\m{}_\s ]T^{\s\n}\fr{1}{\square}\,,
\end{equation}
where we see that the apparent pole $-{1}/({\partial^2_N +\square})$ has
been replaced by an expected massless propagator ${ 1/\square}$. This
should not be understood as a result of the specific gravitational
coupling, but as a characteristic of Feynman computations for the PST
model.
 
The expression of the effective propagator together with the
interaction terms in $S^g_\P$ can further be used to determine the
building blocks of the one-loop Feynman diagrams for the coupling of
the PST model to a gravitational background. A similar method was
carried out in~\cite{lech} in computing the gravitational anomalies in
$4n+2$ dimensions.

\section{Conclusions}\label{s:concl}

In the present paper we demonstrated the equivalence of the PST and
Schwarz-Sen formulations of duality-symmetric Maxwell theory at the
quantum level. The latter, Schwarz-Sen, is quantum mechanically also
equivalent to the ordinary Maxwell theory~\cite{ggrs} such that all
these models are physically related (on-shell) at the classical and
quantum level, even if their off-shell descriptions are different.

To this end (to prove the equivalence) we have adopted the
antifield-BRST quantization method. This approach resides in
compensating all the gauge symmetries of the original system by some
fermionic ghosts and their antibracket conjugates - called
antifields. After extending the action to a suitable chosen
non-minimal sector, we had to fix the gauge. We were able to perform
two different gauge-fixings. The covariant one preserves Lorentz
invariance but it has the disadvantage of an intricate form in the
ghost sector which makes its integration difficult.  On the other
hand, giving up Lorentz symmetry we presented also a non-covariant
gauge which has a simple structure in its fermionic part leading us to
favourable results. Firstly, it was the cornerstone in proving the
quantum equivalence of the studied PST model and the Schwarz-Sen
action. Secondly, the correct Feynman rules have been infered. We used
the example of gravitational interaction of the PST system in the same
gauge condition to show explicitly that the unphysical pole of the
propagator is a gauge artifact, of the same kind that the one
appearing for usual Maxwell theory in Coulomb gauge. Such a
first-order expansion in the perturbed metric was performed also by
Lechner~\cite{lech} in studying the gravitational anomalies of the
self-dual tensors in $4n+2$ dimensions. However, as discussed
in~\cite{agw}, the self-dual vector field in $4$ dimensions should be
anomaly free.

\acknowledgments

We thank K.~Van~Hoof and M.~Henneaux for their encouragements and
support in this project.  We are also grateful to G.~Barnich,
C.~Bizdadea, N.~Boulanger, F.~Roose, S.O.~Saliu, C.~Schomblond,
W.~Troost and A.~Van~Proeyen for useful discussions.  We acknowledge
the organizers of the school ``Supergravity, Superstring and M-Theory"
(september 18th 2000--february 9th 2001) held at the ``Centre \'Emile
Borel" (UMS 839 - CNRS/UPMC), for their hospitality during the last
part of this paper.

\appendix

\section{Basic ingredients of antifield-BRST formalism}\label{a:BRST}

Here we give only some of the main ideas underlying the lagrangian
BRST method. For more details we refer the reader to~\cite{bv,ht}.

Let $S_0[\phi^i]$ be an action with the following bosonic gauge
transformations\footnote{We use the DeWitt notation.}
\begin{equation}
\label{e:GT}
\delta_\varepsilon \phi^i = R^i_\alpha\epsilon^\alpha
\end{equation}
which are irreducible.
Then, one has to enlarge the ``field'' content to
\begin{equation}
\{\Phi^A\}=\{\phi^i,C^{\alpha}\}\,.
\end{equation}
The fermionic ghosts $C^{\alpha}$ correspond to the parameters
$\varepsilon^\a$ of the gauge transformations~(\ref{e:GT}).  To each
field $\Phi^A$ we associate an antifield $\Phi_A^*$ of opposite
parity. The set of associated antifields is then
\begin{equation}
\{\Phi_A^*\}=\{\phi^{*}_i,C^*_\alpha\}\,.
\end{equation}
The fields possess a vanishing antighost number (antigh) and a
nonvanishing pure\-ghost number (pgh)
\begin{equation}
\hbox{pgh}(\phi^i)=0,\qquad \hbox{pgh}(C^\alpha)=1.
\end{equation}
The pgh number of the antifields vanish but their respective
antigh number is equal to
\begin{eqnarray}
\hbox{antigh}(\Phi_A^*)=1+\hbox{pgh}(\Phi^A).
\end{eqnarray}
The total ghost number (gh) equals the difference between the pgh
number and the antigh number.  The antibracket of two functionals
$X[\Phi^A,\Phi_A^*]$ and $Y[\Phi^A,\Phi_A^*]$ is defined as
\begin{equation}
(X,Y)=\int d^nx\left( \frac{\delta^RX}{\delta\Phi^A(x)}
\frac{\delta^LY}{\delta\Phi_A^*(x)}
-\frac{\delta^RX}{\delta\Phi_A^*(x)}\frac{\delta^LY}{\delta\Phi^A(x)}\right)\,,
\end{equation}
where $\delta^R/\delta Z(x)$ and $\delta^L/\delta Z(x)$ denote
functional right- and left-derivatives.

The \emph{extended action} $S$ is defined by adding to the classical
action $S_0$ terms containing the antifields in such a way that the
classical \emph{master equation},
\begin{equation}
(S,S)=0\,,
\label{e:master}
\end{equation}
is satisfied, with the following boundary condition:
\begin{equation}
S=S_0+\phi^*_iR^i_\alpha C^{\alpha}+\dots
\end{equation}
This imposes the value of terms quadratic in ghosts and
antifields. The extended action has also to be of vanishing gh number.
If the algebra is non-abelian, we know that we have to add other
pieces of antigh number two in the extended action with the general
form (due to structure functions)
\begin{equation}
S_2^{2a}=\frac{1}{2}C_\alpha^* f_{\beta \gamma}^\alpha C^\beta
C^\gamma\,.
\end{equation}
If the algebra is open, other terms in antigh number must be added,
quadratic in $\phi^*_i$'s.  Furthermore, other terms in higher antigh
number could be necessary, e.g.\ when the structure functions depend on
the fields $\phi^i$.

The extended action captures all the information about the gauge
structure of the theory: the N\"{o}ether identities, the (on-shell)
closure of the gauge transformations and the higher order gauge
identities are contained in the master equation.

The BRST transformation $s$ in the antifield formalism is a canonical
transformation, i.e.\ $sA=(A,S)$.  It is a differential: $s^2=0$, its
nilpotency being equivalent to the master
equation~(\ref{e:master}). The BRST differential decomposes according
to the antigh number as
$$
s=\d + \gamma + \hbox{"more"}
$$
and provides the gauge invariant functions on the stationary surface,
through its cohomology group at gh number zero $H_0(s)$.  The
Koszul-Tate differential $\delta$
\begin{equation}
\label {e:KT}
\d \F^*_i=(\F^*_i,S)\vert_{\F_A^*=0}
\end{equation}
implements the restriction on the stationary surface, and the exterior
derivative along the gauge orbits $\gamma$
$$\gamma\F^i=(\F^i,S)\vert_{\F^*_A=0}$$
picks out the gauge-invariant functions.

The solution $S$ of the master equation possesses gauge invariance,
and thus, cannot be used directly in a path integral.  There is one
gauge symmetry for each field-antifield pair. The standard procedure
to get rid of these gauge degrees of freedom is to use the
\emph{gauged-fixed action} $S_\P$ defined by
\begin{equation}
S_\P=S_{\rm non-min}\left[ \F^A,\F^*_A=\fr{\d\P[\F^A]}{\d\F^A}\right].
\label{e:GAUGEFIXED}
\end{equation}
The functional $\P[\F^A]$ is known as the \emph{gauge-fixing fermion}
and must be such that $S_\P[\Phi]$ is non-degenerate, i.e.\  the
equations of motion derived from the gauge-fixed action $\d
S_\P[\F^A]/\d\F^A=0$ have unique solution for arbitrary initial
conditions, which means that all gauge degrees of freedom have been
eliminated.  It also has to be local in order that the antifields are
given by local functions of the fields.

The generating functional of the theory is
\begin{equation}
\label{e:PATHINT}
Z=\int[\cD\F^A] \exp iS_\P \,.
\end{equation}
The value of the path integral is independent of the choice of the
gauge-fixing fermion $\P$.  The notation $[\cD\Phi]$ stands for
$\cD\Phi\,\mu [\Phi]$, where $\mu[\F]$ is the measure of the path
integral.  It is important to notice that the expression of the
measure $\mu[\Phi]$ in this path integral is not completely determined
by the Lagrangian approach.  A correct way to determine it, would be
to start from the Hamiltonian approach for which the choice of measure
is trivial, indeed it is known to be $\cD\F\cD\Pi$, that is the
product over time of the Liouville measure $d\Phi^Ad\Pi_A$.

It can be proved that, if correctly handled, the two approaches are
equivalent (see~\cite{ht} and references therein).  This justifies a
posteriori the choice of the measure $\mu [\Phi]$
in~(\ref{e:PATHINT}).

\section{Gauge-fixing of Maxwell theory}\label{a:Max}

The generating functional for the Maxwell theory in the Hamiltonian
approach is well known (see e.g.~\cite{ggrs,ht})
\begin{equation}
Z= \int {\cal D} A_m \,{\cal D}\pi_m \,{\cal D}c \,{\cal D} 
\bar {\cal P}\,{\cal D}\bar c\, {\cal D}{\cal P} \, \exp iS^M_\P\,.
\end{equation}
As usual, $\pi_m$ is the conjugate momentum of $A_m$, $c$ and $\bar c$
are ghosts, and $\bar {\cal P}$ and ${\cal P}$ are their respective
conjugate momenta.  The Hamiltonian gauge-fixed action in the Coulomb
gauge is given by
\begin{equation}
S^M_\Psi= \int d^4 x(\pi_m\dot A^m + \dot c \bar {\cal P} - {\cal H}_0
+ A_0 \partial_i \pi^i + \pi_0\partial^i A_i + i \bar {\cal P}{\cal P}
- i\bar c \square c)\,,
\end{equation}
where $i$, $j$, $\dots$ stand for spatial indices in the 3 dimensional
hyperplane $x^0$ constant.  The Hamiltonian density is equal to
\begin{equation}
{\cal H}_0 = \frac{1}{2}  (\pi^i\pi_i + B^i B_i)\,.
\end{equation}
The magnetic field is $B^i=F^*_{0i}$.  We can easily integrate the
fields $A^0$, $\pi_0$, the ghosts $c$, $\bar c$ as well as their
conjugate momenta ${\cal P}$, $\bar {\cal P}$.  Then, we obtain that
\begin{equation}
Z= \int {\cal D} A_i \,{\cal D}\pi_i \,{\rm
det}(\square)\,\delta(\partial_i \pi^i)\,\delta(\partial^i A_i) \, \exp
i{\tilde S}^M_\P
\end{equation}
with
\begin{equation}
{\tilde S}^M_\Psi= \int d^4 x(\pi_i\dot A^i - {\cal H}_0)\,.
\end{equation}
The determinant of $\square$ comes from the integration on the fermionic
ghosts.  The integration on $A^0$ and $\pi_0$ gives the
delta-functions enforcing, respectively, the Gauss law and the Coulomb
gauge.

In order to make the connection with the gauge-fixed Schwarz-Sen
action, we have to move to a two-potential formulation, that is we
have to solve the Gauss constraint $\partial_i \pi^i=0$ by introducing
a potential $Z^i$ such that
\begin{equation}
\pi^i=\epsilon^{ijk}\partial_jZ_k\,.
\label{e:pi}
\end{equation}
The potential $Z_i$ can be decomposed into a sum of a longitudinal and
a transverse part: $Z_i=Z_i^L+Z_i^T$.  When $Z_i$ is transverse
($Z_i=Z^T_i$), the equation~(\ref{e:pi}) is invertible (with
appropriate boundary conditions).  More precisely, in that case one
expresses $Z_i$ as
\begin{equation}
Z_i=-\bigtriangleup^{-1}\epsilon_{ijk}\partial^j \pi^k\,.
\label{e:zed}
\end{equation}
We can introduce the field $Z^i$ in the path integral in the following
way
\begin{equation}
Z= \int {\cal D} A_i \,{\cal D}\pi_i \,{\cal D}Z_i\, {\rm
det}(\square)\,\delta(\partial_i \pi^i)\,\delta(\partial^i
A_i)\,\delta(Z^i+\bigtriangleup^{-1}\epsilon^{ijk}\partial_j \pi_k) \,
\exp i{\tilde S}^M_\P\,.
\end{equation}
In order to make the comparison with the Schwarz-Sen approach we will
use the relation
\begin{equation}
\delta(Z^i+\bigtriangleup^{-1}\epsilon^{ijk}\partial_j \pi_k) =
\delta(Z^{L\,i})\delta(Z^{T\,i}+\bigtriangleup^{-1}\epsilon^{ijk}\partial_j
\pi_k)
\end{equation}
with $\delta(Z^{L\,i})=\delta(\partial_iZ^i)$. We also notice that
\begin{eqnarray}
&&\delta(Z^{T\,i}+\bigtriangleup^{-1}\epsilon^{ijk}\partial_j
\pi^T_k)\,=\,\underbrace{{\rm det}^{-1}\,(\bigtriangleup^{-1}{\rm
curl})}_{={\rm det}({\rm
curl})}\,\delta(\pi^{T\,i}-\epsilon^{ijk}\partial_jZ^T_k)\,,
\end{eqnarray}
where ``${\rm curl}$" stands for the operator
$\epsilon^{ijk}\partial_j\,,$ and $\partial_i\pi^{T\,i}=0$.  We
finally identify the two potentials as follows
\begin{equation}
A^{1}_{i} =  A_i \,, \qquad A^{2}_{i} =  Z_i\,. 
\end{equation}
Putting all these remarks together we can integrate out the $\pi_i$ to obtain
\begin{equation}
\label{e:genfunc}
Z= \int {\cal D} A^\a_i \, {\rm det}(\square)\,{\rm det}({\rm
curl})\,\delta(\partial^i A^\a_i) \, \exp iS^{S-S}_\P\,,
\end{equation}
where $S^{S-S}_\P$ is the Schwarz-Sen gauge-fixed action
\begin{equation}\label{e:ScSe}
S^{S-S}_\Psi= \int d^4 x\,\fr12 (\cL^{\a\b}\dot{A}^\a_i
-\delta^{\a\b}B^\a_i)B^{\b\, i}\,.
\end{equation}

\end{document}